\documentclass[numberedappendix,apj]{emulateapj}
\usepackage{amssymb,amsmath}
\usepackage{natbib}
\usepackage{graphicx}
\shorttitle{Mira's cometary head/tail}
\shortauthors{Esquivel et al.}
\slugcomment{Draft version, \today}

\begin{document}

\title{A model of Mira's cometary head/tail entering the Local
Bubble}

\author{
A. Esquivel\altaffilmark{1},
A. C. Raga\altaffilmark{1}, J. Cant\'o\altaffilmark{2},
A. Rodr\'{\i}guez-Gonz\'alez\altaffilmark{1},\\
D. L\'opez-C\'amara\altaffilmark{1},
P. F. Vel\'azquez\altaffilmark{1}, and F. De Colle\altaffilmark{3}}
\affil{
\altaffilmark{1}Instituto de Ciencias Nucleares, Universidad Nacional
Aut\'{o}noma de M\'{e}xico, Apartado Postal 70-543, 04510 M\'{e}xico
D.F., M\'{e}xico\\
\altaffilmark{2}Instituto de Astronom\'\i a, Universidad
Nacional Aut\'onoma de  M\'exico, Ap. 70-468, 04510 D.F., M\'exico\\
\altaffilmark{3}Department of Astronomy and Astrophysics,
University of California Santa Cruz, Santa Cruz, CA 95064, USA
} 

\email{esquivel, raga ,ary, d.lopez, pablo@nucleares.unam.mx;
fabio@ucolick.org}

\begin{abstract}

We model the cometary structure around Mira as the interaction of an
AGB wind from Mira A, and a streaming environment. Our simulations
introduce the following new element: we assume that after
${200~\mathrm{kyr}}$ of evolution in a dense environment Mira entered
the Local Bubble (low density coronal gas).
As Mira enters the bubble, the head of the comet expands
quite rapidly, while the tail remains well collimated for a
${>100~\mathrm{kyr}}$  timescale. The result is a
broad-head/narrow-tail structure that resembles the
observed morphology of Mira's comet.
The simulations were carried out with our new adaptive
grid code {\sc Walicxe}, which is described in detail.
\end{abstract}


\section{Introduction}
\label{sec:introduction}
\begin{figure*}
\centering
\includegraphics[width=.9\textwidth]{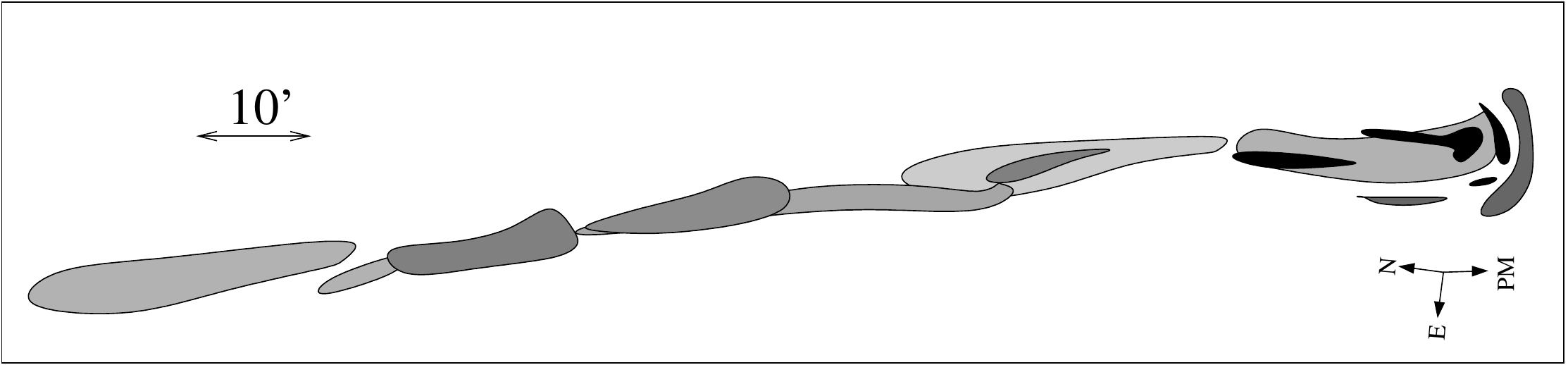}
\caption{Sketch of the UV image of the tail of Mira observed with {\it
    GALEX}, fig.\,1a of \citet{2007Natur.448..780M}. The intensity is
  represented by the different shades of gray (darker corresponds to a
  higher intensity). The orientation (N,E) and proper motion (PM) are
  indicated by the arrows below the head of the comet.}
\label{fig:tail}
\end{figure*}

Mira is a well studied binary
system. Mira A is a prototypical, thermally pulsating luminous AGB star
(which in fact gives name to the so-called Mira class of variables),
while Mira B is a much less luminous star, believed to be a
white dwarf or a main sequence star. Recently, the discovery of
a cometary structure of $\sim 2\arcdeg$ in the sky with its head
centered on Mira, by \citet{2007Natur.448..780M}, with the {\it GALEX}
satellite, has generated a renewed interest in this system.
In order to compare the observations with our models we include in
Figure \ref{fig:tail} a schematic diagram of the UV image of
\citet{2007Natur.448..780M}.

The broad-band UV filters (FWHM $\Delta\lambda=256,~730\,$\AA, centered at 
$\lambda_c=1516,~2267\,$\AA)  in {\it GALEX} make difficult to
discriminate the physical mechanism responsible for the observed
emission.
\citet{2007Natur.448..780M} suggested that most of the observed
emission in the tail of Mira could be due to fluorescent $\mathrm{H_2}$
lines, while in regions closer to the bow-shock it is likely that
other species, such as \ion{C}{4} could contribute to the emission.

The distance to Mira, estimated from {\it Hipparcos} is
$107~\mathrm{pc}$ \citep{2003A&A...403..993K}. 
The system Mira moves at a high speed ($\sim 
125~\mathrm{km~s^{-1}}$) with respect to its surrounding medium, as
deduced from its proper motion \citep{1993BICDS..43....5T} and radial
velocity \citep{1967IAUS...30...57E}.
The physical size of the cometary tail, assuming the {\it Hipparcos}
distance is $\sim 4\mathrm{pc}$.

New observations and/or reexamination of previous observations have
been made in several wavelenghts, including \ion{H}{1} 21cm
\citep{2008ApJ...684..603M}, IR \citep{2008ApJ...687L..33U}, and
optical \citep{2009A&A...500..827M}.
Recent theoretical work has been done by
\citet{2007ApJ...670L.125W, 2008ApJ...680L..45R, 2008ApJ...685L.141R}.
\citet{2007ApJ...670L.125W} performed a 3D hydrodynamical simulation
of the interaction between an isotropic AGB wind and the ISM, modeled as a
plane-parallel wind (considering a reference system
in which Mira is at rest). \citet{2008ApJ...680L..45R} included an
inhomogeneous AGB wind, resulting in a more complicated shock structure
for the cometary head (possibly more consistent with the observations). 
These two models, however, have one flagrant flaw: if the models
are adjusted to produce the correct size for the cometary head,
the tail that they produce is significantly broader than the one
of Mira's comet. 

The reason for this discrepancy with the observations lies
in the fact that the ISM density in those models is estimated
by assuming ram pressure balance between Mira's wind and the
ISM  at the tip of the head of the cometary structure. Therefore, the
distance between the front of the bow-shock and Mira is consistent
with the observations. However, the tail produced by the model has
a width comparable to the head, while Mira's comet shows a
considerably narrower tail (see Fig \ref{fig:tail}).

\citet{2007ApJ...670L.125W} speculated that
Mira could have recently crossed into the Local Bubble (a tenuous
region of coronal gas). Most of the tail could then have formed
outside the Local Bubble, in a higher density ISM, resulting in
the observed, narrow tail structure. In the present paper, we
explore this possibility through a set of axisymmetric numerical
simulations of a travelling star+wind system which emerges into
a hot, coronal gas bubble.

The paper is organized as follows: in section 2 we describe the model,
including a review of the analytical solution of the wind/streaming
environment by
\citet{1996ApJ...459L..31W} and \citet{1996ApJ...469..729C}. The
setup of the numerical simulations is presented in section 3, followed
by the results in section 4, and a summary in section 5. As this paper
is the first publication with our new hydrodynamical code {\sc
  Walicxe}, we include an appendix describing the code in detail, as
well as some tests.

\section{The model}
\label{sec:model}
\begin{figure}
\centering
\includegraphics[width=0.4\textwidth]{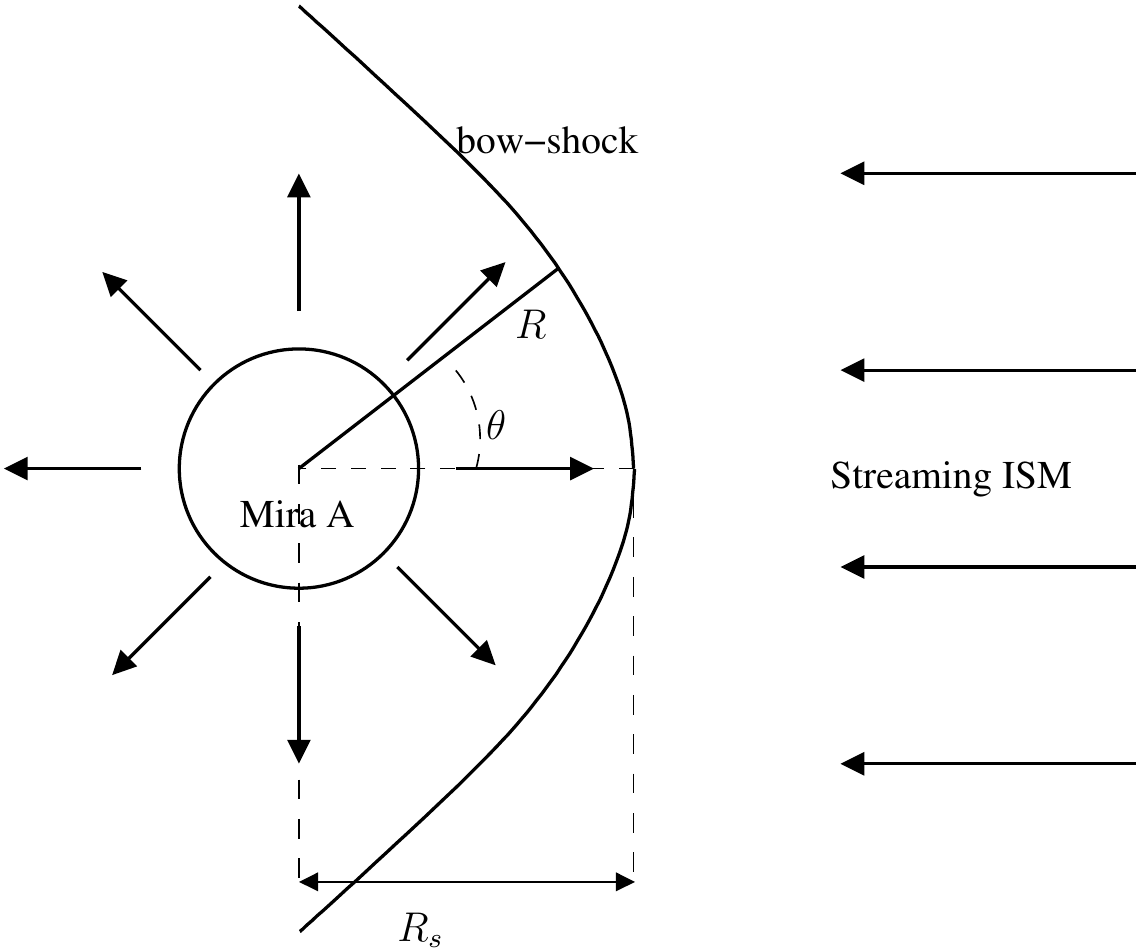}
\caption{Schematic of the two-wind interaction model. The interaction
  of a moving spherical wind source (Mira A) and its surrounding ISM
  can be modeled, in a reference frame with Mira A at rest, as a
  stationary wind source accompanied by a streaming ISM. 
  In this figure, we also illustrate the spherical radius $R$ used in
  the analytical model, the polar angle $\theta$, and the standoff
  distance $R_{\mathrm{s}}$.}
\label{fig:diagram}
\end{figure}

Following \citet{2007ApJ...670L.125W} and \citet{2008ApJ...680L..45R} we
model Mira's cometary tail as the interaction of a slow, dense,
wind that arises from the AGB star Mira A, with a fast,
lower-density, plane-parallel wind that results from the motion of Mira
through the ISM (the simulations are performed in the reference frame
in which Mira is at rest), as shown in
the schematic diagram of Figure \ref{fig:diagram}.

The structure that results from the interaction between a
streaming, plane-parallel flow and a spherical wind
has a cometary shape that has
been described analytically, considering conservation of linear and
angular momentum in a ``thin-shell'' approximation
\citep{1996ApJ...459L..31W,1996ApJ...469..729C}.
With this formalism, the shape of bow-shock is found to have the form:
\begin{equation}
R(\theta)=R_{\mathrm{s}}\csc\theta\sqrt{3(1-\theta\cot\theta)},
\label{eq:comet}
\end{equation}
where $R$ is the spherical radius (measured from the center of the
isotropic wind source), $\theta$ is the polar angle (measured from
the symmetry axis, aligned with the direction towards the impinging,
plane-parallel flow), and $R_{\mathrm{s}}$ is the 
stagnation radius (or standoff distance, see Figure \ref{fig:diagram}). 
The stagnation radius (point of closest approach to the spherical wind
source) is given by the ram-pressure balance between the two winds:
\begin{equation}
\frac{\dot{M}_{\mathrm{w}}v_{\mathrm{w}}}{4\pi R_{\mathrm{s}}^2}=\rho_{\mathrm{a}}v_{\mathrm{a}}^2,
\label{eq:Rs}
\end{equation}
where $\dot{M}_{\mathrm{w}}$ is the mass loss rate of Mira,
$v_{\mathrm{w}}$ its terminal wind velocity, $\rho_{\mathrm{a}}$ the
ambient (ISM) mass density, and $v_{\mathrm{a}}$ the velocity
of the star moving through the surrounding environment.
\begin{figure}
\centering
\includegraphics[width=0.5\textwidth]{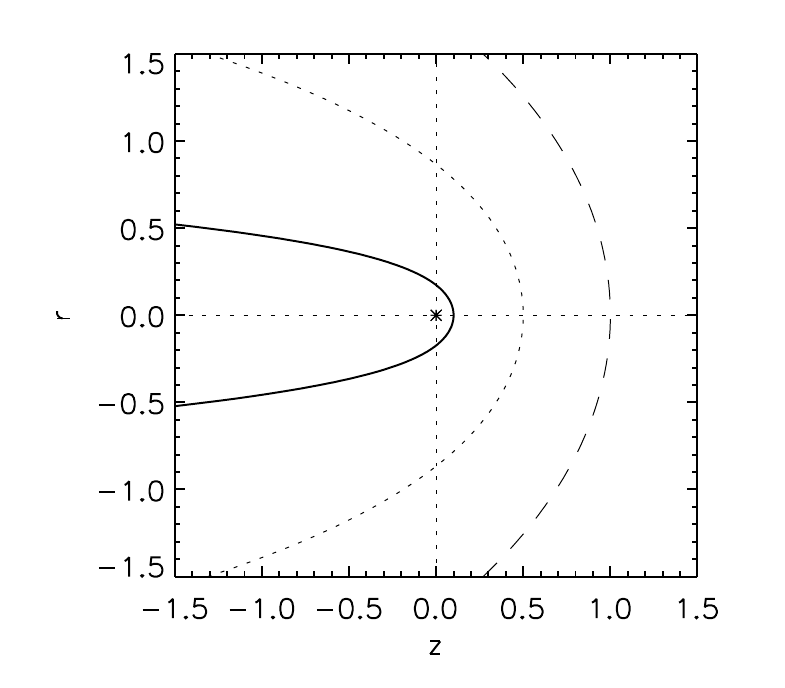}
\caption{Analytical solution of the two-wind ``thin-shell''
  model \citep{1996ApJ...459L..31W,1996ApJ...469..729C} for three
  different stagnation radii:
${R_{\mathrm{s}}=0.1,~0.5, 1.0}$ ({\it solid}, {\it
    dotted} and {\it dashed} lines, respectively). The position of the
  spherical wind source is at $(0,0)$, marked by the star. The
  plane-parallel wind impinges form the right as in Fig.
  \ref{fig:diagram}.}
\label{fig:model}
\end{figure}

Figure \ref{fig:model} shows the locus of the interaction region
(location of the thin-shell, and shape of the bow-shock structure) for
different stagnation radii. Notice that for smaller $R_{\mathrm{s}}$
the cometary tail becomes narrower.

\section{The numerical setup}
\label{sec:setup}

We performed two hydrodynamical simulations of Mira's tail with the
new code {\sc Walicxe}. The code solves the gas-dynamic equations in a
block based adaptive mesh that is designed to run in parallel on machines
with distributed memory (i.e. clusters). 

{\sc Walicxe} can be run with a Cartesian two-dimensional mesh, or an
axisymmetric (cylindrical) grid. 
For this particular problem we have used the axisymmetric version of
the code with the hybrid HLL-HLLC Riemann solver (see the Appendix for
details).

Along with the gas-dynamic equations, a rate equation for hydrogen is
solved. The resulting H ionization fraction is used to
compute the radiative energy losses, which
are obtained with the parametrized cooling function (that depends on
the temperature, density and hydrogen ionization fraction) described in
\citet{2004RMxAA..40...15R}.

The simulations have 5 root blocks of $16\times16$ cells each, aligned
along the axial direction ($z-$axis). We allow $7$ levels of
refinement, yielding an equivalent resolution of $1024\times 5120$
(radial$\times$axial) cells at the highest grid resolution. The computational
domain extends $[(0,0.25)\times(0,1.25)]\times 10^{19}$ cm
in the radial and axial
directions, respectively. Thus, the resolution at the highest level of
refinement is $\approx 2.4\times 10^{15}$ cm.
We impose a reflective boundary condition along the symmetry axis 
($r=0$). Open boundary conditions are used at $r=0.25\times
10^{19}~\mathrm{cm}$ and $z=0$.

In both models the spherical wind has the following properties.
It is injected (at every timestep) in a
spherical region of $3\times 10^{16}\mathrm{cm}$ radius, centered
at $r=0$, $z=1.05\times 10^{19}~\mathrm{cm}$. The density inside the
injection region follows an $\propto R^{-2}$ law ($R$ is the spherical
radius measured from the position of the wind source), scaled so that
the mass-loss rate is $\dot{M}_{\mathrm{w}}=3\times
10^{-7}~\mathrm{M_{\odot}yr^{-1}}$. The wind is neutral, has a temperature of
$T_{\mathrm{w}}=10^2~\mathrm{K}$, and an outward, radially directed velocity
$v_{\mathrm{w}}=5~\mathrm{km~s^{-1}}$. The values of the physical
parameters of Mira's wind were taken from the observational studies of
CO lines by \citet{1995ApJ...445..872Y} and  \citet{2000ApJ...545..945R}.

At the initial time ($t=0$), both models are identical, with a
plane-parallel wind that fills the computational domain 
(except inside the sphere in which the isotropic wind is imposed), which is
replenished by an inflow condition at $z=1.25\times 10^{19}~\mathrm{cm}$. 
The ISM is homogeneous, with a hydrogen number density
$n_{\mathrm{H}}=1~\mathrm{cm^{-3}}$ (consistent with the ISM in the
galactic plane, just outside the Local Bubble in the vicinity of Mira,
see \citealt{2001A&A...366.1016V}), and a velocity 
$v_{\mathrm{a}}=125~\mathrm{km~s^{-1}}$ (in the $-z$ direction).
The wind and ISM are
neutral, except for a seed electron density assumed to arise from
singly ionized C (i.e. the minimum ionization fraction is $10^{-4}$).
The ISM has a temperature of $T_{\mathrm{a}}=10^3~K$.

In model M1, we evolve these initial/boundary conditions
up to an integration time of $350~\mathrm{kyr}$. Model M1 corresponds
to the evolution of Mira outside of the Local Bubble, in a uniform
medium dense enough to produce the narrow tail as observed.

In model M2, we simulate Mira entering the Local Bubble, a region much
hotter and tenuous than the average ISM.
To achieve this, after $200~\mathrm{kyr}$ of evolution in the same
medium as model M1, we raise the temperature of the ISM that enters
the computational domain to $T_{\mathrm{a}}=10^6~K$, lower its density
to $n_{\mathrm{H}}=5\times 10^{-2}~\mathrm{cm^{-3}}$ \citep[from the
observations by][]{2003A&A...411..447L}, and change its
ionization fraction to $1$ (fully ionized hydrogen). We then let this
configuration evolve up to a $t=350~\mathrm{kyr}$ integration time.

The model of Mira entering the Local Bubble is well justified by
the current Galactic Position of Mira $(l,b)=(168,-58)$
\citep{1997A&A...323L..49P, 2007ApJ...670L.125W}. This position puts
Mira inside, but close to the edge of the Local Bubble, as can be seen
from the maps presented by \cite{2003A&A...411..447L}. Given the
high velocity of Mira, in the $\sim 300~\mathrm{kyr}$ that it
takes to form a tail of $4~\mathrm{pc}$, it would have traveled
approximately $38~\mathrm{pc}$. This distance is enough to place the
starting point of the tail formation outside the Local Bubble.

\section{Results}
\label{sec:results}
\begin{figure*}
\centering
\includegraphics[width=\textwidth]{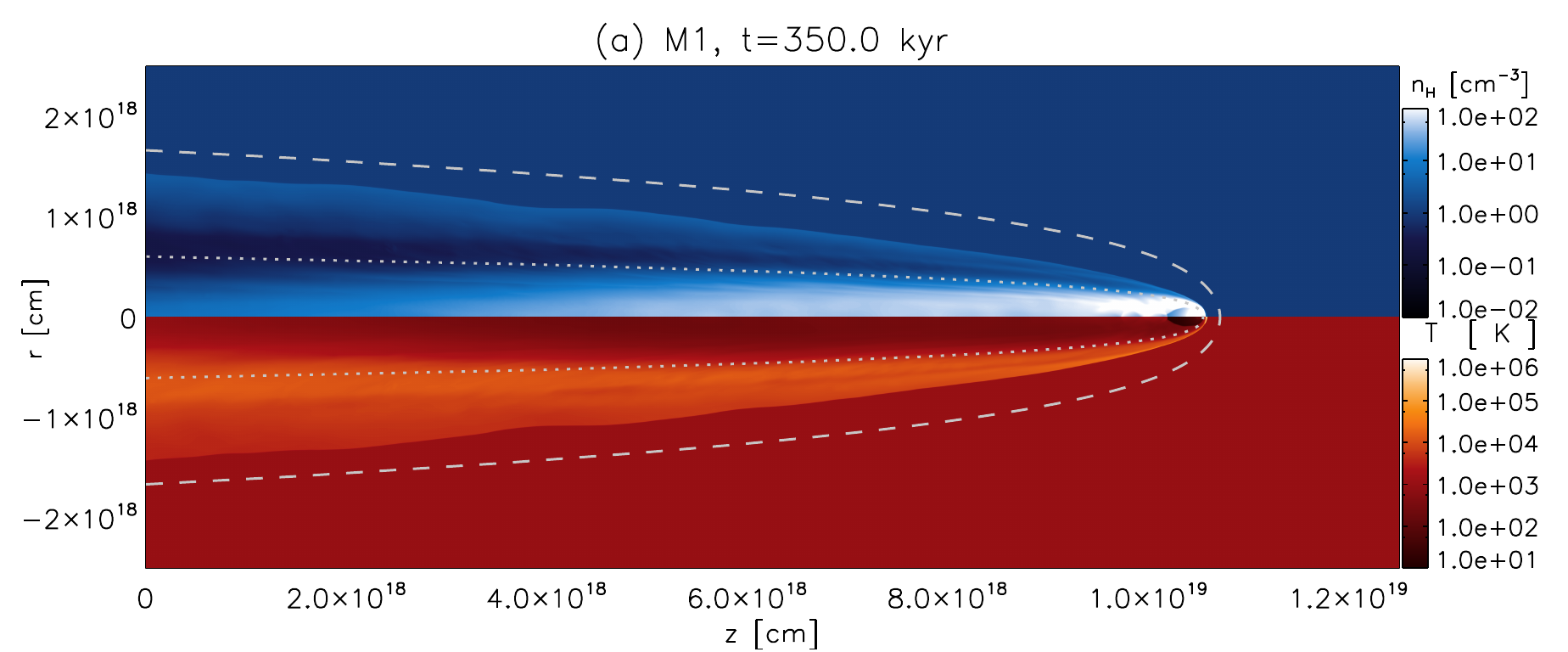}
\includegraphics[width=\textwidth]{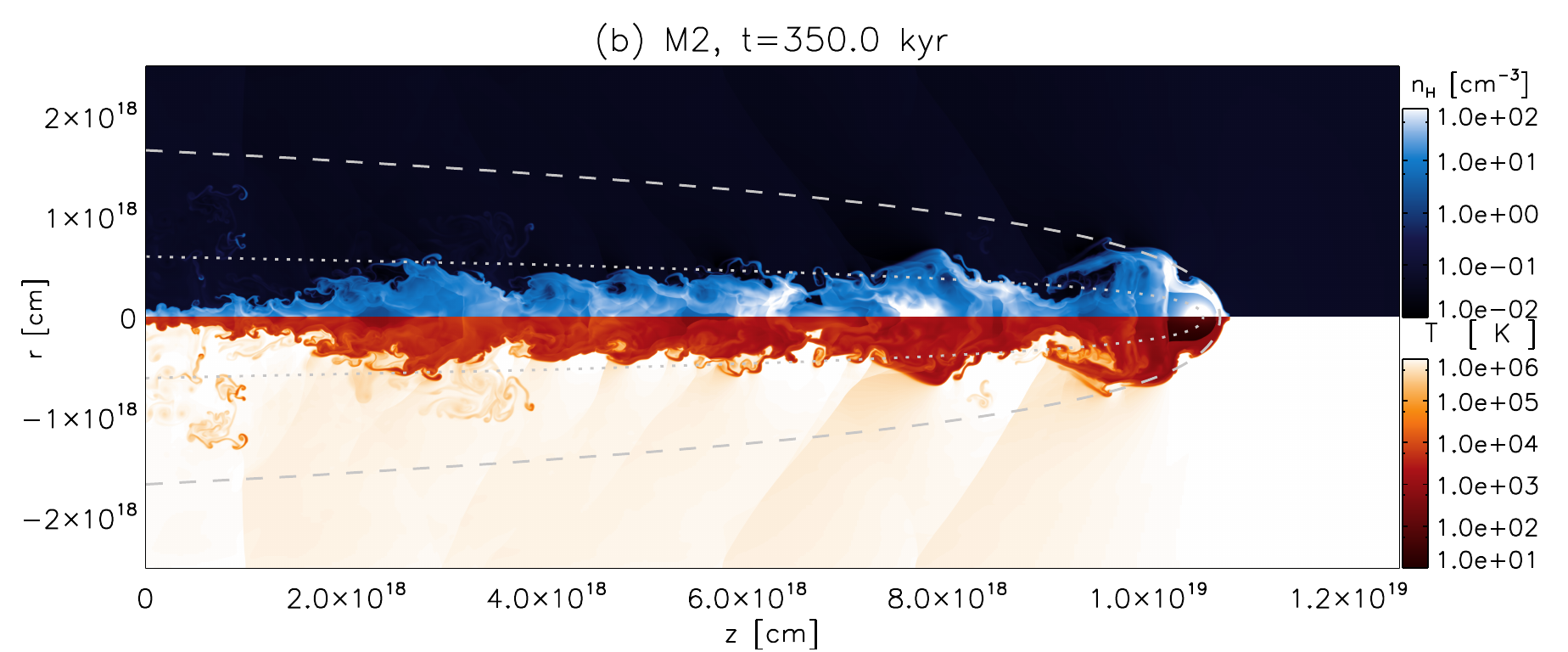}
\caption{Density (top half of each panel), and temperature (bottom
  half) stratifications at a
  $t=350~\mathrm{kyr}$ integration time, for the two models.
  The numerical values are shown in the logarithmic gray-scale
  (color-scale in the on-line version) bars at the right of each panel.
  We have included in both panels the analytical solution for the
  bow-shock for the two stagnation radii used (see text for details):
  the {\it dotted} line is for $R_{\mathrm{s}}=4.9\times
  10^{16}~\mathrm{cm}$ and the dashed line for $R_{\mathrm{s}}=2.2\times
  10^{17}~\mathrm{cm}$.
  The temperature panel has been reflected on the symmetry axis
  for visual purposes.} 
\label{fig:rhot}
\end{figure*}

We let the models to evolve up to an integration time of
$350~\mathrm{kyr}$, which is of the order of the time needed to form
a cometary tail of $\sim 4~\mathrm{pc}$ in length (the size of the
structure observed with {\it GALEX}).
Figure \ref{fig:rhot} shows the density (top half of the panels)
and temperature distributions (bottom half) of M1 and M2 at the final
evolutionary stage.

From the top panel in Figure \ref{fig:rhot} we can see that after
$350~\mathrm{kyr}$ a narrow tail, of axial and radial
dimensions comparable with the
observations of Mira, is formed. However, the stand-off distance (in
agreement with 
the analytical prediction in eq. \ref{eq:Rs}) is  $R_{\mathrm{s}}\sim
4.9\times 10^{16}~\mathrm{cm}$, which is much closer to the
source compared to the $R_{\mathrm{s}}\sim 3.1\times
10^{17}~\mathrm{cm}$ that can be deduced from the observations
of Mira's comet. The
analytical solution for the parameters of model M1, plotted as a
dotted line, traces well the region that separates the dense and cold
material which arises from Mira's wind, and the hotter, lower
density shocked ISM. 
This can be seen more clearly in the following section, where we
separate the contribution of the wind and show column density maps.

The bottom panel of Figure  \ref{fig:rhot} shows the tail that would
be produced if Mira entered the Local Bubble some $145~\mathrm{kyr}$
ago (i.~e., model M2).
The analytical solution for the parameters used for the Local Bubble
is included as a dashed line. The corresponding standoff distance for
model M2 is at $R_{\mathrm{s}}=2.2\times 10^{17}~\mathrm{cm}$ to the
right of Mira, a distance that is close to what is observed
($3.1\times 10^{17}~\mathrm{cm}$).

This standoff distance, and the corresponding width of the tail
(dashed line) are much larger than observations. However, the change
from one stationary configuration to the other is not
instantaneous. As the cometary
flow expands after entering the Local Bubble the head grows rapidly.
Approximately $20~\mathrm{kyr}$ after the entrance of Mira into the
Local Bubble we obtain a structure that resembles remarkably well the
broad-head/narrow-tail of Mira's comet, such tail remains well
collimated for at least $100~\mathrm{kyr}$. It is only after
$150~\mathrm{kyr}$ of entering the Local Bubble that the tail becomes
somewhat wider than the UV observations (see time sequence in the 
following section).

A clear difference in the wind/environment interface can be seen
in models M1 and M2. In model M2, the outer boundary of the region
filled by the stellar wind shows a more complex structure, which
fragments the tail downstream from the wind source.
Such structures are absent in Model M1
(see Figure \ref{fig:rhot}). This difference is a result of the fact
that while the wind/environment boundary has a highly supersonic
velocity shear in model M1, this shear is barely transonic
in model M2 (as Mira's motion is basically sonic with
respect to the hot, Local Bubble environment). It is a well
known experimental result that the spreading rate of sonic
mixing layers is much larger than the one of high Mach number
mixing layers \citep[see][]{1991ApJ...372..646C}, this effect
is also present in our models, therefore a more turbulent tail is seen
in model M2.

\subsection{Column density maps}
\begin{figure*}
\centering
\includegraphics[width=0.5\textwidth]{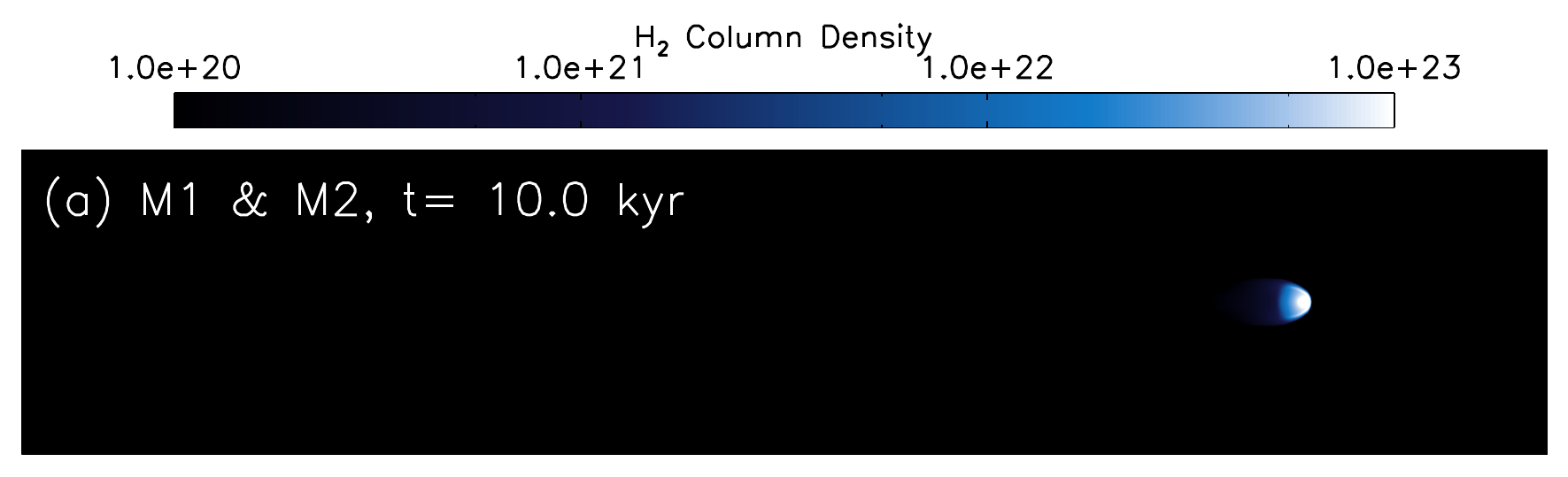}\\
\includegraphics[width=0.5\textwidth]{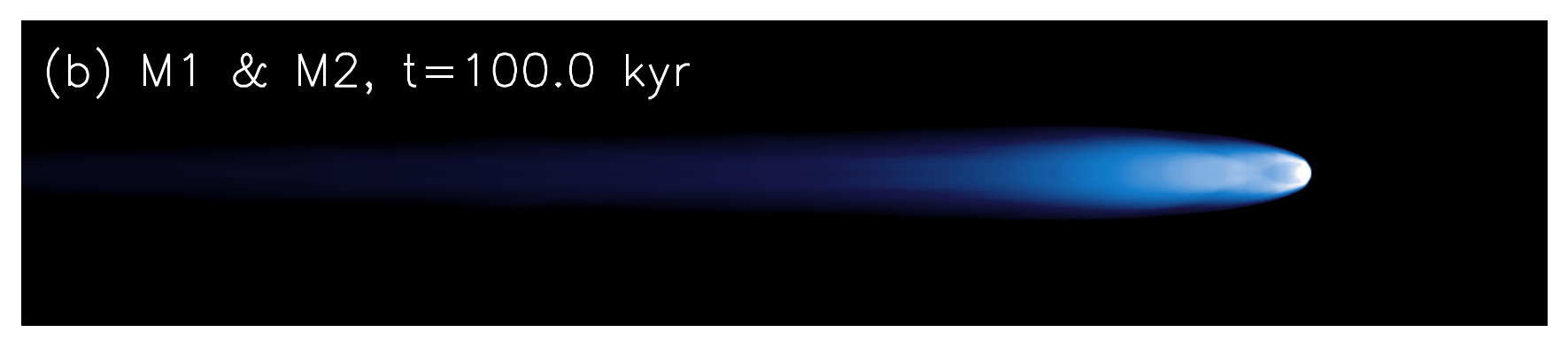}\\
\includegraphics[width=0.5\textwidth]{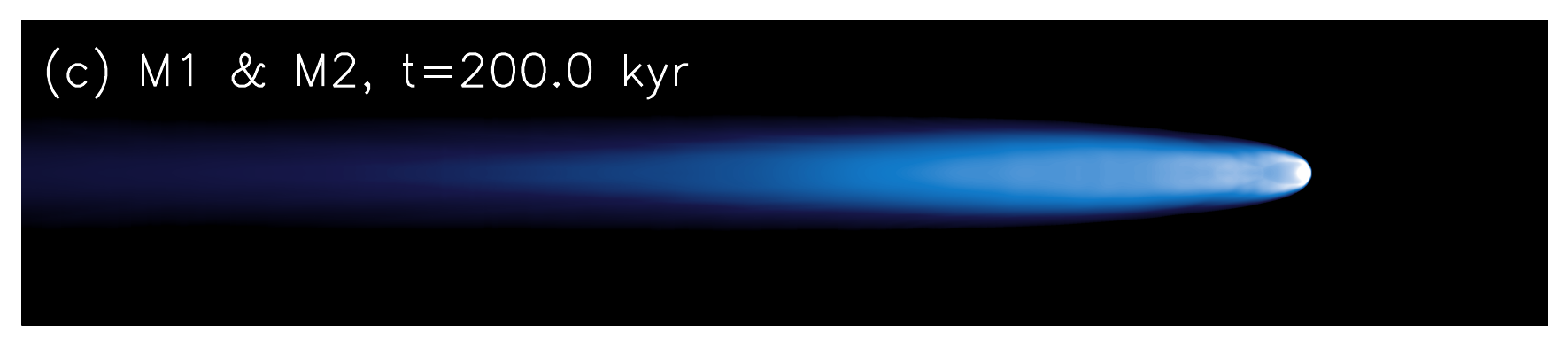}\\
\includegraphics[width=0.5\textwidth]{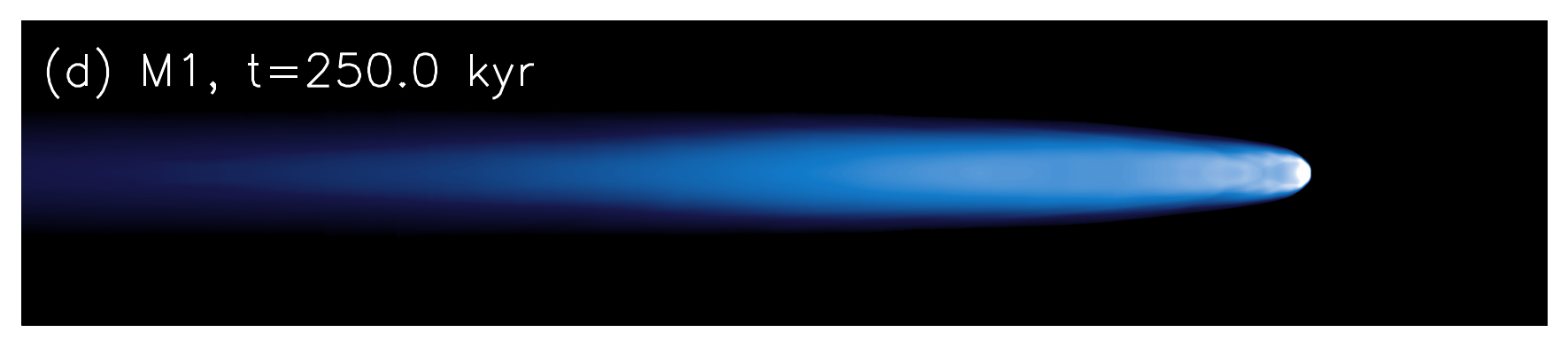}\includegraphics[width=0.5\textwidth]{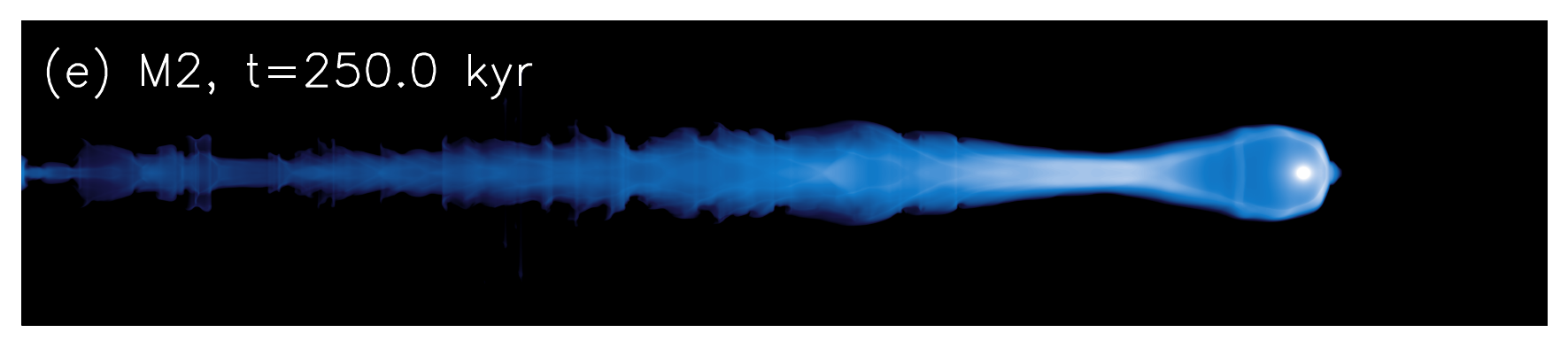}\\
\includegraphics[width=0.5\textwidth]{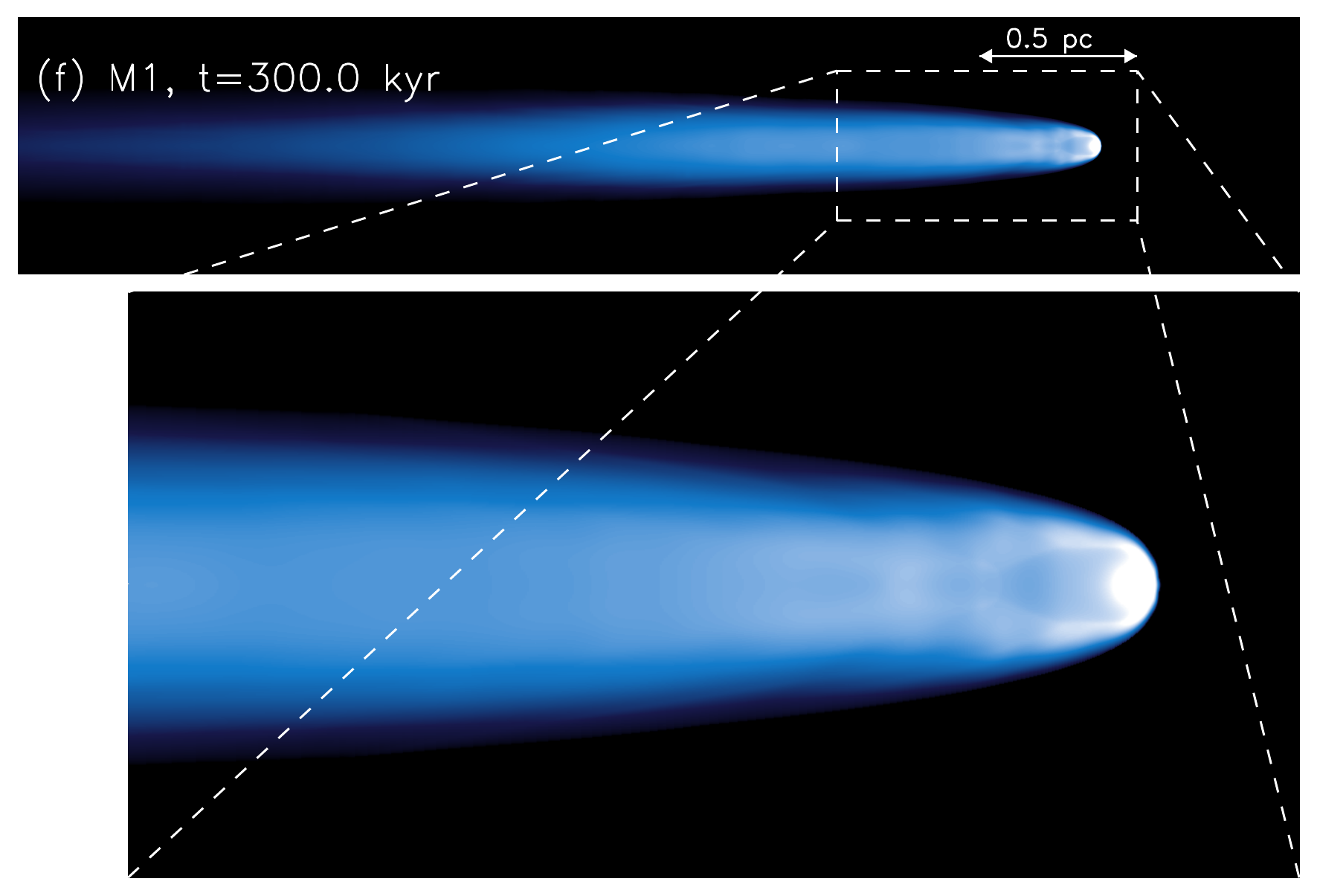}\includegraphics[width=0.5\textwidth]{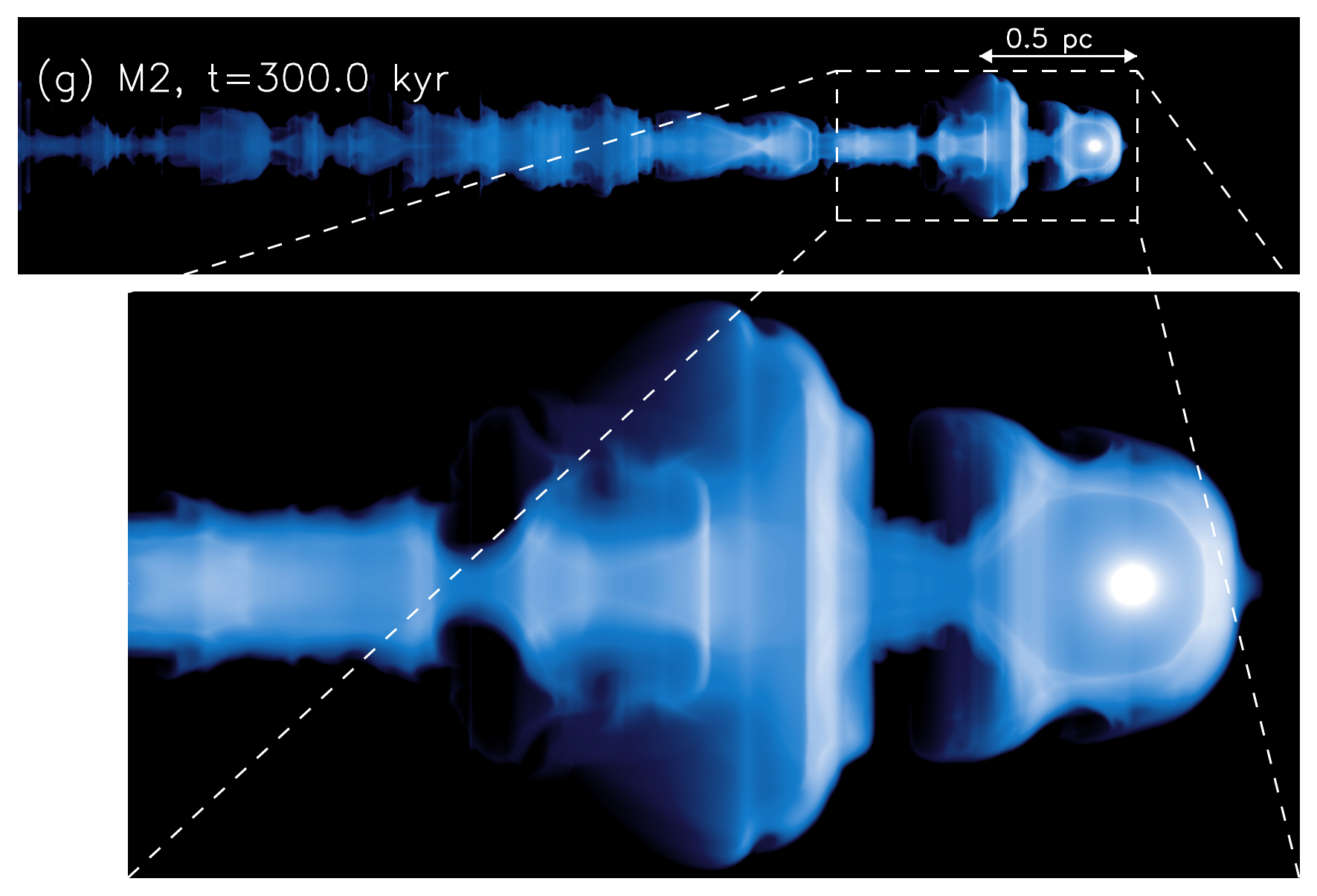}\\
\includegraphics[width=0.5\textwidth]{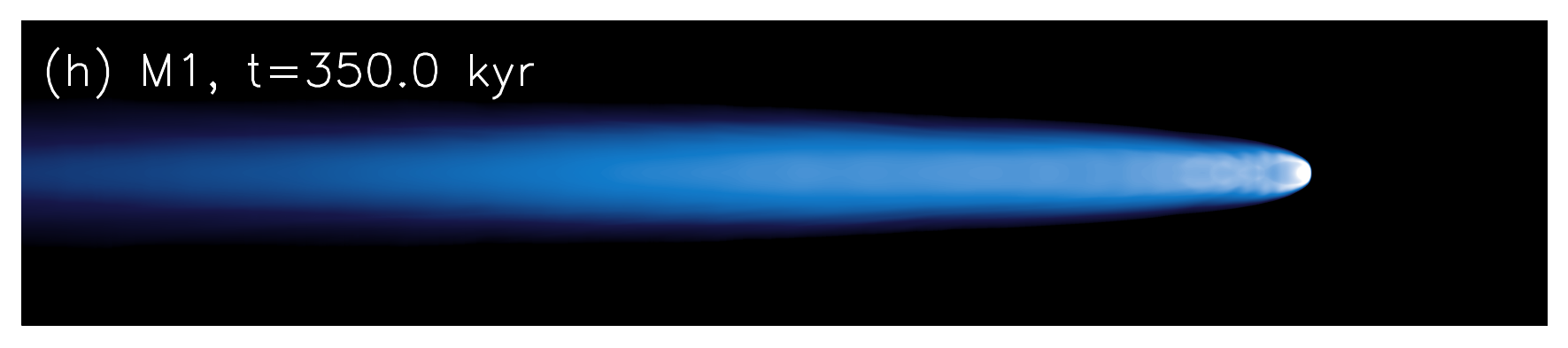}\includegraphics[width=0.5\textwidth]{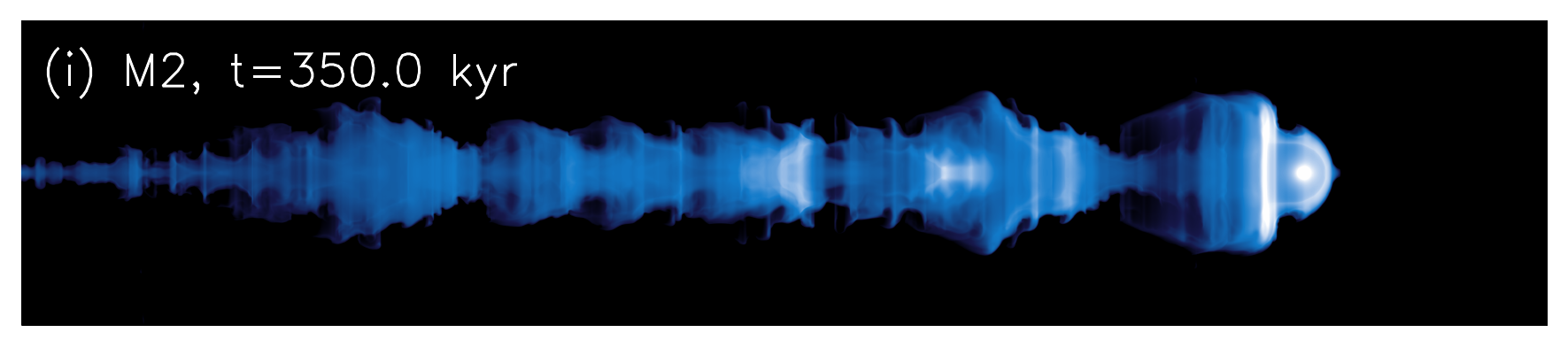}\\

\caption{Time sequence of H$_2$ column density
  (i.~e., of the material from Mira's wind),
  for models M1 and M2. The value of the
  column density can be read from the logarithmic gray-scale bar
  (color-scale in the on-line version) at the top of the plots. The
  time of evolution is indicated in the legend at the upper left
  corner of each panel. For the first three panels ({a}--{\it d}) the
  evolution is identical for both models. We have included a blow-up
  of the region close to the wind source for an integration time of 
  $300~\mathrm{kyr}$ ({\it f} and {\it g}). The physical size is
  indicated by the arrows in panels ({\it f}) and ({\it g}).}
\label{fig:column}
\end{figure*}

One problem of interpreting the \citet{2007Natur.448..780M} observations
of Mira is that the broad band filters in {\it GALEX} do not
give much information about the physical mechanisms that produce the
observed emission. They suggested that the bulk of the emission arises
from H$_2$ molecules, which are likely to be
present in the cold wind of Mira, but not in the surrounding
environment.

In order to make a better comparison with the observations we have
added a passive scalar in the simulations that allows us to identify the
material from Mira's wind and the ISM material. We then
computed column density maps of the material from Mira's wind
(which would be molecular H$_2$) and
present them in the form of a time sequence in Figure
\ref{fig:column}.

In Figure \ref{fig:column} we see that
the tail forms and reaches the observed length (of Mira's comet)
after approximately $\sim 300~\mathrm{kyr}$. The ISM that
enters from the right changes (in model M2) to the Local Bubble parameters at
$t=200~\mathrm{kyr}$, but it takes some $\sim 5~\mathrm{kyr}$ to reach
the two-wind interaction region. From then on, models M1 and M2 start to
differ.

From the column density maps it is clear that the stellar wind
material in M1 is well confined to a region that resembles the tail of
Mira for the duration of the simulation, but (as we have discussed
above) this requires an ambient density that puts the axial standoff
distance too close to Mira. 

For model M2, as Mira crosses into
the Local Bubble its wind expands towards a new steady
state configuration (which is not reached in our simulations).
We found that at after only $20\mathrm{kyr}$ the stellar wind
has expanded enough to form a dense structure extending ${\sim 2\times 
  10^{17}~\mathrm{cm}}$ upstream from the stellar wind
source. This size is in agreement with the axial extend of the head of
Mira's comet. 
We also see that the tail in model M2 expands at a much slower rate
compared to the head, therefore reproducing the ``broad head/narrow
tail'' morphology of Mira's comet for a timescale on the order of
$100~\mathrm{kyr}$. After $150~\mathrm{kyr}$ inside the Local Bubble the
tail of Mira has expanded laterally beyond what is observed in the
{\it GALEX} UV maps (see \citealt{2007Natur.448..780M}, and diagram in
Fig. \ref{fig:tail}). 

To simulate the UV emission that would arise from the H$_2$ molecules
requires a chemical/radiation transfer calculation that is beyond
the scope of this paper. We only note that if we use the
standard conversion factor $A_{\mathrm{V}}\approx
10^{-21}N_{\mathrm{H}}$ (due to dust absorption), we
find that the tail in our simulations has $A_{\mathrm{V}}\sim 10$.
Therefore, molecules present in the wind from Mira A would not
be photodissociated in a substantial way, and the
tail should indeed have a large H$_2$ fraction.

Two interesting issues arise from this work, one is the age
of Mira's tail itself, the other is how long it has been inside the
Local Bubble (assuming that this is the case).

Regarding the age of the tail, in the original discovery of
\citet{2007Natur.448..780M}, the authors estimate that the tail was
formed in approximately $30~\mathrm{kyr}$, while the simulations of
\citet{2007ApJ...670L.125W} suggested an age of
$450~\mathrm{kyr}$. This large discrepancy is related to the rate at
which Mira's wind decelerates to merge dynamically with the
ISM. \citet{2007Natur.448..780M} assumed that the wind material
instantly decelerates, so the $30~\mathrm{kyr}$ is the time that Mira
takes to traverse the length of the tail. In contrast, the
$450~\mathrm{kyr}$ timescale of \citet{2007ApJ...670L.125W} was
directly obtained from numerical simulations which are, however, sensitive to
viscous effects. Such effects depend mainly on two things: the density
of the ISM that is used (a physical effect), and the amount of
artificial viscosity (a numerical effect) needed to stabilize the
code.
Artificial viscosity usually overwhelms real viscosity, resulting in
an enhanced deceleration, thus the numerical estimates correspond to
lower limits of the actual age. 
Our simulations lie in between these two previous estimates, as they
require a time of $\sim 300~\mathrm{kyr}$ to develop Mira's tail. This
difference with the model of Waering et al. (2007) is due to the
higher ISM density of our models, which results in a higher
deceleration rate. 

The question of the time spent by Mira inside the Local Bubble is
not well constrained by our models because the tail takes on the
order of $100~\mathrm{kyr}$ to expand laterally in an appreciable way.
Our results suggest that Mira could have crossed into the Local Bubble
between $20$ and $120~\mathrm{kyr}$ ago. This result is in rough
agreement with the $130$-$225~\mathrm{kyr}$ proposed by
\citet{2007ApJ...670L.125W}.
In this respect, a more detailed map of the Local Bubble and an
extrapolation of Mira's motion might prove more useful.

\section{Summary}
\label{sec:summary}

We have modeled the cometary structure observed around Mira as the
interaction of the cold, dense wind from Mira A with a streaming
ISM. For this purpose we have used a newly developed code, which is
described in detail in the Appendix.
Similar simulations have been done previously
\citep{2007ApJ...670L.125W, 2008ApJ...680L..45R}, but they fail to
reproduce the broad-head/narrow-tail configuration characteristic of
Mira's comet.

The new element in our simulations is to assume that
Mira has recently crossed (from a denser than previously considered)
medium into the Local Bubble, a region of tenuous coronal gas. This
possibility was first suggested by \citet{2007ApJ...670L.125W}, and
is quite feasible given Mira's current location.

By starting a stellar wind in a denser region than previous papers we are
able to reproduce the width of the observed tail of Mira's comet.
By allowing the stellar wind source to enter (at a later time) a
less dense region of the ISM (i.~e., the Local Bubble)
we obtain an expansion of the cometary head. For the
parameters of our model, the predicted cometary structure
has a striking resemblance to Mira's comet after $\sim 20~\mathrm{kyr}$
time after entering the Local Bubble, and the structure last for
another $\sim 100~\mathrm{kyr}$.

The age that we obtain for the formation of the tail is $\mathrm{\sim
  300~\mathrm{kyr}}$. This can be compared with estimates by
\citet[][$30~\mathrm{kyr}$]{2007Natur.448..780M} and by
\citet[][$\mathrm{450~\mathrm{kyr}}$]{2007ApJ...670L.125W}, that assume
either total, or very little (respectively) deceleration of the wind
as it encounters the ISM. 

We should note that \citet{2009A&A...500..827M} recently discovered
a bipolar jet system in the Mira system (possibly ejected from
Mira B). This collimated outflow might have an important effect
in the formation of Mira's cometary head and tail, which has not
been considered in our present work, and should clearly be
studied in the future.

\acknowledgments
This work has been supported by CONACYT grants 61547, 101356, and
101975.
We thank Enrique Palacios, Mart\'{\i}n Cruz and Antonio Ram\'{\i}rez
for maintaining the cluster in which the calculations of this paper
were carried out.

\appendix
\section{The {\sc walicxe} code} 
\label{sec:appendix}

Our previous adaptive grid codes (e.~g., the
{\sc yguaz\'u-a}, see \citealt{2000RMxAA..36...67R})
were not suited for running in parallel computers with distributed memory
(i.e. clusters). In order to achieve high resolution by today's
standards we have developed the new code {\sc walicxe},\footnote{The
  name is one of the few remaining words of the language of the
  charr\'uas, a South-American tribe than inhabited what is now
  Uruguay and part of Argentina and Brazil. The meaning is
  ``sorcery'', which fits to some extent the job description of
  theoretical/numerical astrophysicists.}
which contains the cooling/chemistry modules of {\sc yguaz\'u-a}, but
uses a new adaptive mesh designed to be easily parallelizable in clusters.
At this point we only have a 2D version, with a 3D version in
progress.
We describe below the numerical algorithm and the AMR scheme of the
{\sc walicxe} code.

\subsection{Basic equations}
The 2D Euler system of
equations can be written in Cartesian coordinates as:
\begin{equation}
\frac{\partial \mathbf{U}}{\partial t}
+ \frac{\partial \mathbf{F}}{\partial x}
+ \frac{\partial \mathbf{G}}{\partial y}
=\mathbf{S},
\label{eq:Euler}
\end{equation}
where
\begin{equation}
    \mathbf{U} =\left[ \begin{array}{c}
        \rho \\ \rho u\\\rho v\\  E \\ n_1 \\ n_2 \\ \vdots \\ n_r
      \end{array}\right],~~
    \mathbf{F}=\left[\begin{array}{c}
        \rho u \\P+ \rho u^2\\\rho u v\\ u(E+P) \\ n_1u \\ n_2u \\
        \vdots \\ n_ru 
      \end{array}\right],~~
    \mathbf{G}=\left[\begin{array}{c}
      \rho v \\\rho v u \\ P+ \rho v^2\\ v(E+P) \\ n_1v \\ n_2v \\
      \vdots \\ n_rv
    \end{array}\right],~~\mathrm{and}~~
  \mathbf{S}=\left[\begin{array}{c}
      0 \\ 0 \\ 0 \\ G-L \\ S_1 \\ S_2 \\ \vdots \\ S_r
    \end{array}\right].
    \label{eq:U}
\end{equation}
$\mathbf{U}$ are the so-called conserved variables, $\rho$ is the mass density,
$(u,v)$ are the velocity components in the $(x,y)$-directions,
$n_1,n_2,..., n_r$ are densities of additional species that are
advected passively into the fluid, which can be used to compute
heating/cooling rates. $\mathbf{F}$ and ${\mathbf{G}}$ are the fluxes
in the $x$ and $y$ directions, respectively;
\begin{equation}
  P=\left( n_i+n_e \right){k T}
\end{equation}
is the thermal pressure, and the (total) energy is given by
\begin{equation}
  E=\frac{1}{2}\rho\left(u^2+v^2\right)+C_v P,
\end{equation}
where $C_v$ is the specific heat at constant volume.
$\mathbf{S}$ is the source vector, that contains the energy gains and
losses ($G$ and $L$, from radiative and collisional processes) and the
reactions rates for the $n_1, n_2,...n_r$ species, 
the source vector can be also extended to include geometrical terms
for cylindrical and spherical coordinate systems.
In addition, one can define a vector with the so-called primitive
variables
\begin{equation}
    \mathbf{W} =\left[ \begin{array}{c}
        \rho \\  u \\ v\\  P \\ n_1 \\ n_2 \\ \vdots \\ n_r
      \end{array}\right], 
\end{equation}
which are easily obtained from the conserved variables.

\subsection{Discretization: finite-areas}

We will denote the indices and grid spacings of the computational
cells in the  $x,~y$-directions by $i,~j$ and $\Delta x,~\Delta y$,
respectively. The time is discretized with an index $n$ such that 
$\Delta t_n=t_{n+1}-t_n$, which in principle varies at each time-step,
however to make the notation clearer we will drop the sub-index $n$
from $\Delta t_n$. 

Integration of equation \ref{eq:Euler} on the area of the
computational cell centered at $i~,j$, on a time interval
$[t+\Delta t]$ yields the following {\it exact} expression:
\begin{equation}
\mathbf{U}_{i,j}^{n+1}=
\mathbf{U}_{i,j}^n
-\frac{\Delta t}{\Delta x}\left(\mathbf{F}_{i+1/2,j}^{n+1/2} 
                               - \mathbf{F}_{i-1/2,j}^{n+1/2} \right)
-\frac{\Delta t}{\Delta y}\left(\mathbf{G}_{i,j+1/2}^{n+1/2} 
                              -  \mathbf{G}_{i,j-1/2}^{n+1/2}\right)
+\Delta t \mathbf{S}_{i,j}^{n+1/2},
\label{eq:exact-sol}
\end{equation}
where the variables $\mathbf{U}_{i,j}^n$ are now area averages at a
time $t=t_n$: 
\begin{equation}
\mathbf{U}_{i,j}^n=\frac{1}{\Delta x \Delta y}
\int_{y-1/2}^{y+1/2} \int_{x-1/2}^{x+1/2}
\mathbf{U}\left(x,y,t_n \right) \,dx\,dy,
\label{eq:Uij}
\end{equation}
where $x\pm 1/2=x_i\pm \Delta x/2$, $y\pm 1/2=y_i \pm \Delta y/2$ are
the locus of the boundaries of the cell at $x_i,~y_j$. The fluxes
are now time (and length) averages:
\begin{equation}
\mathbf{F}_{i+1/2,j}^{n+1/2}=\frac{1}{\Delta y \Delta t}
\int_{t_n}^{t_{n+1}}\int_{y-1/2}^{y+1/2}
\mathbf{F}\left(x_{i+1/2},y,t \right) \,dy\,dt,
\label{eq:Fij}
\end{equation}
\begin{equation}
\mathbf{G}_{i,j+1/2}^{n+1/2}=\frac{1}{\Delta x \Delta t}
\int_{t_n}^{t_{n+1}}\int_{x-1/2}^{x+1/2}
\mathbf{G}\left(x,y_{j+1/2},t \right) \,dx\,dt .
\label{eq:Gij}
\end{equation}And the source terms are time and area averages:
\begin{equation}
\mathbf{S}_{i,j}^{n+1/2}=\frac{1}{\Delta x \Delta y \Delta t}
\int_{t_n}^{t_{n+1}}\int_{y-1/2}^{y+1/2}\int_{x-1/2}^{x+1/2}
\mathbf{S}\left(x,y,t \right) \,dx\,dy\,dt .
\label{eq:Sij}
\end{equation}
The method we have described if often referred to as {\it
  finite-volumes}, but since we are dealing with 2D it is
more accurate to call it {\it finite-areas}.
Of course, the expressions of eqs. (\ref{eq:exact-sol})--(\ref{eq:Sij})
are easily extended to three dimensions yielding a truly finite-volumes
method, where the area integrals used above become volume integrals,
and the length integrals become surface integrals.
What remains now, is to find approximations to the numerical intercell fluxes
($\mathbf{F}_{i+1/2,j}^{n+1/2},~\mathbf{G}_{i,j+1/2}^{n+1/2} $) and
sources ($\mathbf{S}_{i,j}^{n+1/2}$). 
This can be done by solving the Riemann problem at the cell
interfaces, that is, assuming a true discontinuity of the hydrodynamic
variables at such interfaces \citep{torobook}.

\subsection{Solution method: Finite areas, second order Godunov scheme} 

We use a second order Gudunov type method to advance the solution,
it uses a linear reconstruction (with slope limiters) of the primitive
variables and an approximate Riemann solver to compute the fluxes at
the intercell boundaries. As it is customary, below we
explain the method in one dimension, the expressions for the fluxes in
the $y$-direction are analogous to those in the $x$-direction.
The finite-volume (finite-lengths in 1D) solution is:
\begin{equation}
\mathbf{U}_{i}^{n+1}=
\mathbf{U}_{i}^n
-\frac{\Delta t}{\Delta x}\left(\mathbf{F}_{i+1/2}^{n+1/2} 
                               - \mathbf{F}_{i-1/2}^{n+1/2} \right)
+\Delta t \mathbf{S}_{i}^{n+1/2},
\label{eq:exact-1d}
\end{equation}
To obtain a second order approximation of the intercell fluxes we
proceed as follows. First, we calculate the timestep $\Delta t$ in
order to ensure that the standard Courant-Friedrichs-Lewy 
condition \citep{CFL} is met. Then, a first order
half-timestep is computed as: 
\begin{equation}
\mathbf{U}_{i}^{n+1/2}=
\mathbf{U}_{i}^n
-\frac{\Delta t}{2 \Delta x}\left(\mathbf{F}_{i}^{n} 
                               - \mathbf{F}_{i-1}^{n} \right)
+\frac{\Delta t}{2} \mathbf{S}_{i}^{n},
\label{eq:1sthalf}
\end{equation}
where the fluxes are obtained solving the Riemann problem with an
approximate Riemann solver \citep[see][]{torobook}, using the
primitive variables 
$\mathbf{W}_i(\mathbf{U}_{i})$. That is
\begin{equation}
\mathbf{F}_i=\mathrm{Riemann}\left(\mathbf{W}_i,\mathbf{W}_{i+1}\right).
\end{equation}
 With the values of
$\mathbf{U}_{i}^{n+1/2}$ from eq. (\ref{eq:1sthalf}) we obtain a new
set of primitives $\mathbf{W}_{i}^{n+1/2}$. We then use a linear
reconstruction to extrapolate them to the cell boundaries
\begin{subequations}
\begin{eqnarray}
\mathbf{W}_{L,i}^{n+1/2}=\mathbf{W}_i^{1/2}-\frac{1}{2}
\mathrm{avg}\left(\mathbf{W}_{i}^{n+1/2}-\mathbf{W}_{i-1}^{n+1/2},
        \mathbf{W}_{i+1}^{n+1/2}-\mathbf{W}_{i}^{n+1/2}
 \right), \\
\mathbf{W}_{R,i}^{n+1/2}=\mathbf{W}_i^{1/2}+\frac{1}{2}
\mathrm{avg}\left(\mathbf{W}_{i}^{n+1/2}-\mathbf{W}_{i-1}^{n+1/2},
        \mathbf{W}_{i+1}^{n+1/2}-\mathbf{W}_{i}^{n+1/2}
 \right),
\end{eqnarray}
\label{eq:qlr}
\end{subequations}
where $\mathrm{avg}$ is an averaging function. The choice of this function is
made in order to limit the slope used and avoid spurious
oscillations. The code has a number of different averaging 
functions, the most widely known (although the most diffusive) is the
$minmod$ function. 

The left and right states in eqs. (\ref{eq:qlr}) are used to approximate
the fluxes used in eq. (\ref{eq:exact-1d}):
\begin{subequations}
\begin{eqnarray}
\mathbf{F}_{i+1/2}^{n+1/2}=
\mathrm{Riemann}\left(\mathbf{W}_{R,i}^{n+1/2}, \mathbf{W}_{L,i+1}^{n+1/2}\right), \\
\mathbf{F}_{i-1/2}^{n+1/2}=
\mathrm{Riemann}\left(\mathbf{W}_{R,i-1}^{n+1/2}, \mathbf{W}_{L,i}^{n+1/2}\right).
\end{eqnarray}
\label{eq:Fapp}
\end{subequations}
At this point a new $\Delta t$ is calculated, and we iterate until the
desired evolutionary stage is reached.
The time step is the same for all blocks, independently of their level
of resolution, this has a cost in computing time, and introduces
additional numerical viscosity to coarser blocks (because they are
effectively run at a smaller Courant number), but it makes easier and
more efficient to balance the load among different processors.

As to the Riemann solver used, the default is the HLLC algorithm, an
improvement proposed by \citet{1994ShWav...4...25T} to the HLL
\citep{HLL} solver, which restores the information of the
contact wave into the solver (the C stands for contact). The original
HLL is also available in the code, but is rarely used because it is
too diffusive. 
In addition to this solvers we have implemented a hybrid scheme that
combines the HLL and HLLC fluxes to fix the so-called carbuncle or
shock instability problem, which occurs when strong shocks align with
the computational grid. Our HLL-HLLC scheme is very similar to those
proposed by \citet{2009JCoPh.228.7634K, 2010IJNMFhuang}, the
description and some tests are given in \S\ref{ap:carb} .

\subsection{The adaptive grid}

The hydrodynamic+rate equations described above are integrated on a
binary, block-based computational grid with the following
characteristics. 

\subsubsection{Blocks (patches)}
The adaptive mesh was designed to be used parallel computers with
distributed memory (i.e. clusters). It consists of a series of
``root'' blocks with a predetermined (set at compile time) number of
cells, $n_x\times n_y$. Each block can be subdivided in a binary
fashion into four blocks with the same number of cells of the parent
block. We will refer to these new blocks as ``siblings'', which
have twice the resolution of their parent block. The maximum number of
levels of refinement allowed ($n_{levs}$) is set at compile time. 
Thus the maximum resolution is
$\Delta_{i}=L_{i}/[2^{n_{levs}-1}(n_{i})]$, where $L_{i}$ is
the size of the computational mesh on the $ith$ direction
(i.e. $i=x,~y$). In addition, each block has its own ghost cells, two
in each direction for the second order scheme described below, but can
be changed to implement higher order schemes. These cells would be
communicated with MPI when adjacent blocks are in different processors.

The  result is a tree structure (of blocks or
patches, not individual 
cells), the root blocks branch into higher resolutions up to
$n_{levs}$. The equations are only solved in ``leaf'' blocks, that
is, only on blocks that are not further refined.

\subsubsection{Refinement/coarsening criteria}

Before advancing the solution, at every timestep we update the mesh.
This is done by sweeping all the leaf blocks, measuring
pressure and density gradients on each of them. If either the density
or pressure suffer a relative gradient, between two adjacent cells,
above a threshold (user defined, typically of $\sim 0.05$) the block
is flagged for refinement. 
In addition, we enforce that neighboring blocks differ at most on one
level of refinement, blocks in proximity to a higher resolution
that has been marked for further refinement are flagged
as well.
Coarsening is also controlled by measuring gradients of density and
pressure, but with a different tolerance (also user defined, typically of
$\sim 1.0$), if all the relative gradients in the block are below this value
the block is marked to be coarsen.
Coarsening is only allowed when an entire group of four siblings have been
marked and it is not impeded by the proximity criterion.

The actual refinement algorithm assigns the same value of the
variables in the parent block to all the siblings ($Oth-$order
interpolation), whereas the coarsening assigns the mean value of each
variable from the four siblings to their parent block.
 
Additional refinement criteria are easy to implement depending of the
problem to be solved. One can, for instance, use a vorticity criteria,
refine above a certain value of any variable, in certain fixed regions, or in
regions labeled  by a passive scalar advected into the flow.

\subsubsection{Parallelization and load balancing}
\label{ap:bal}
Since each of the blocks have the same number of cells, it is
straightforward to send different blocks (actually a group of blocks)
to different processors. This is achieved with the standard Message
Passing Interface (MPI). At every timestep the blocks that share
border with other processors communicate their boundaries (ghost
cells). 
Domain decomposition is made following a Hilbert ordering in
similar fashion to the {\sc Ramses} code \citep{2002A&A...385..337T}.
Each block is assigned a number according to its position in a Hilbert
curve of the  $n_{levs}-1$ order. Following such curve, the number of
leaf blocks is divided  evenly among the available processors, and the
blocks are distributed to be solved. The tree structure is copied
entirely to every processor, but only the leaf blocks are
distributed. 
This, however, is a time consuming operation, and the work-load is only
balanced every $n_{load}$ timesteps (user defined typically
$n_{load}\sim 10-100$), yielding an overhead of about $10$ percent,
depending on how fast the mesh structure changes on a particular 
problem.

\subsection{Testing the code}\label{ap:testing}

Several standard tests have been made to the code, we present here
one that we believe is very illustrative: the classic {\it double Mach
  reflection test} problem proposed by \citet{1984JCoPh..54..115W}.
The test was inspired on laboratory studies and consists of a planar
shock that meets a reflecting surface at an angle of $\pi/3$. The
reflecting surface is taken to be the $x-$axis, and the initial
conditions (explained in all detail in \citealt{1984JCoPh..54..115W})
of the problem are:
\begin{equation}
  \left(\rho, u, v, P \right)=\left\{
\begin{array}{ll}
  \left(1.4,~0,~0,~1 \right) &~~~\mathrm{for}~ x>x_s, \\
  \left(8,~8.25\sin \alpha,~-8.25\cos \alpha,~116.5 \right) &~~~\mathrm{otherwise.}\\
\end{array}\right.
\end{equation}
which correspond to a Mach 10 shock in air (using an adiabatic equation
of state with $C_v=2.5$, $\gamma=1.4$).
The problem is solved on a Cartesian domain
$[0,4]\times[0,1]$ with the following boundary conditions: outflow
boundaries in $x=4$; reflective boundaries in $y=0$ for $x>1/6$; at
$x=0$, and $y=0$ for $x<1/6$ inflow conditions are used; at $y=1$
time dependent inflow conditions are applied, tracing the position of
the shock front that moves according to
$x_s(t)=1/6+10\,y\sin\alpha+y\tan\alpha$.
The flow configuration produces a series of shocks, two of which are
almost perpendicular and are separated by a contact discontinuity, and
a small jet at the base of the reflecting surface. This jet is
driven by a pressure gradient and its evolution is known to be very
sensitive to numerical diffusion. The results of this test can be
compared with those of \citet{1984JCoPh..54..115W}, and many other
codes which have adopted it as a standard test for two dimensional
hydrodynamics (e.g. \citealt{2007ApJS..170..228M,
  2008ApJS..178..137S}).

\begin{figure*}
\centering
\includegraphics[width=0.5\textwidth]{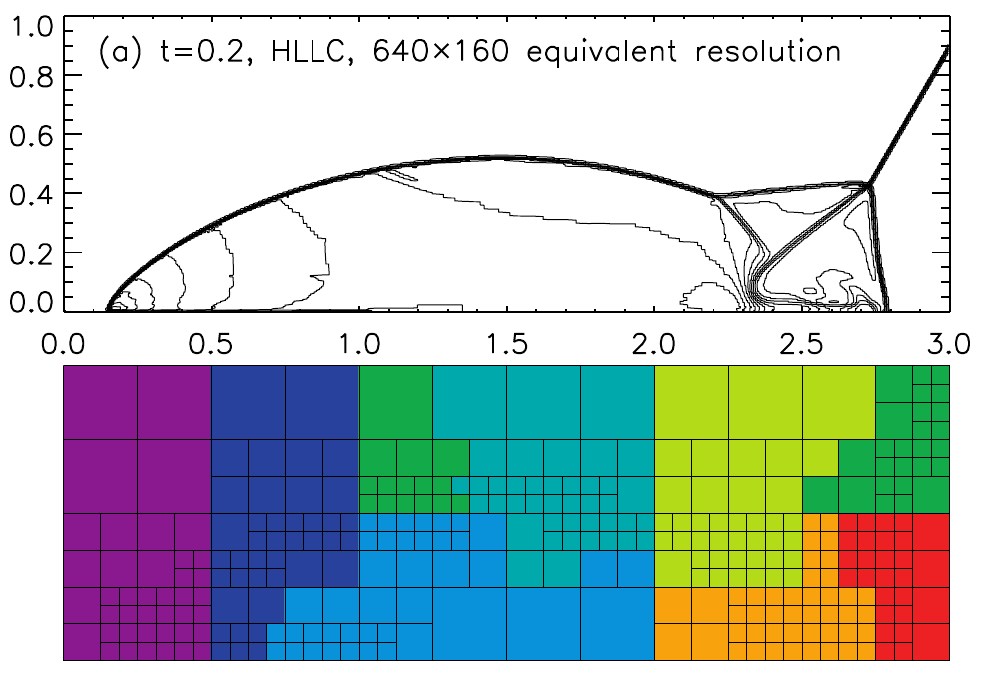}\includegraphics[width=0.5\textwidth]{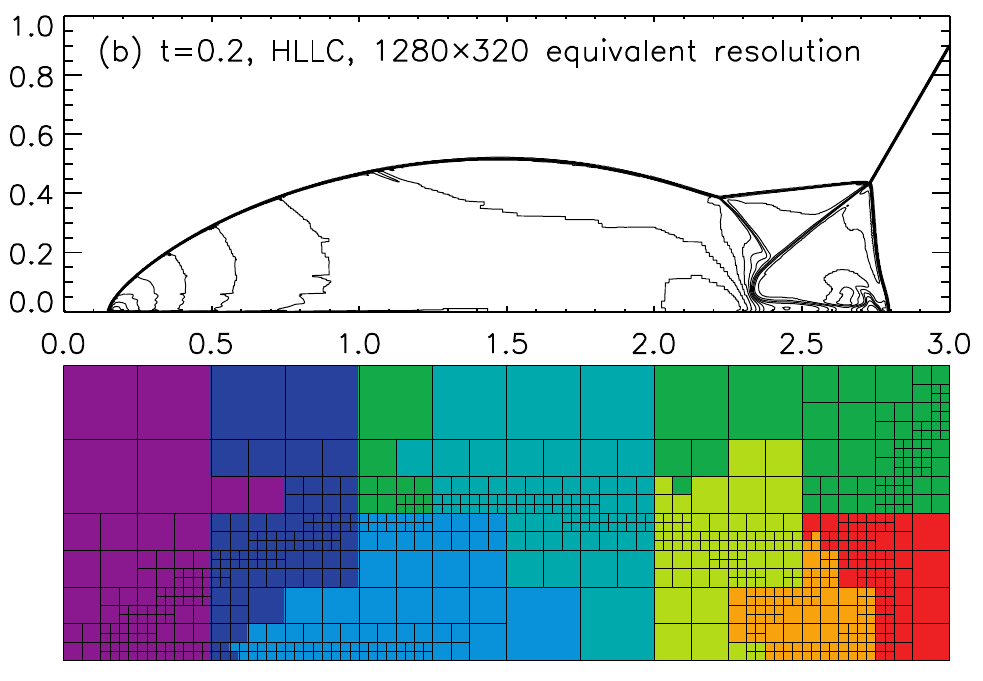}
\caption{Density contours and computational mesh at $t=0.2$ for the
 double Mach reflection test, with the HLLC solver, at two different
 resolutions.
 The computational domain is $[0,4]\times[0,1]$, but for visual purposes
 we only display the $[0,3]\times[0,1]$ region. The equivalent
 resolution at the finest level is indicated in each panel. Below the
 contours we show the AMR blocks (each of $10\times 10$ cells), and with colors
 we indicate the different processors in which the work load was
 distributed at the time.}  
\label{fig:ds}
\end{figure*}

In Figure \ref{fig:ds}, we show the results of the double Mach
reflection test with the HLLC solver, for an integration time of
$t=0.2$, for two different resolutions.
Both runs start with 4 square root blocks aligned along the
$x$-axis. The root blocks have  $10\times 10$ cells, covering a
physical size of $(1 \times 1)$.
The run in panel ({\it a}) has
$5$ levels of refinement, for a
maximum resolution of $\Delta x = \Delta y = 1/160$; the run in
panel ({\it b}) has $6$ levels of refinement, and a maximum resolution of
$\Delta x = \Delta y = 1/320$. Below each of the panels, the AMR
structure (only the $[0,3]\times[0,1]$ region) is shown, each square
corresponds to one $n_x\times n_y$ block, and in color we show the
domain decomposition in $8$ processors. Notice how the mesh is
only refined at places where the flow properties change, and it
remains coarse at regions where the flow is smooth.

Another feature that can be identified is the fact that the jet formed
along the reflecting surface (at $x\sim 2.4$--$2.8$) reaches the
shock that is almost perpendicular to the surface ($x\sim2.8$),
bending it slightly at its base. This is often called ``the
kinked Mach stem'',  and it is unphysical, caused by insufficient
numerical diffusion at places where shocks closely align with the
computational grid \citep{1994IJNMF..18..555Q}. This issue becomes
particularly critical for high resolution simulations (less
numerical/artificial viscosity).

\subsubsection{Parallel performance}
\label{ap:performance}
Is is customary to measure the performance of a parallel code in terms
of its scaling with the number of processors. Ideally the computing
time would scale as $\propto 1/N_p$, where $N_p$ is the number of
processors. In fact, many fixed grid codes show near ideal scaling. 
The reason for such a good scaling is twofold: (i) most of the code is
parallelizable, which restricts the communication between processors
to passing boundaries and determining the timestep, and (ii) since the
grid is fixed, each processor knows beforehand with which 
processors the communication will take place. 

\begin{figure}
\centering
\includegraphics[width=0.5\textwidth]{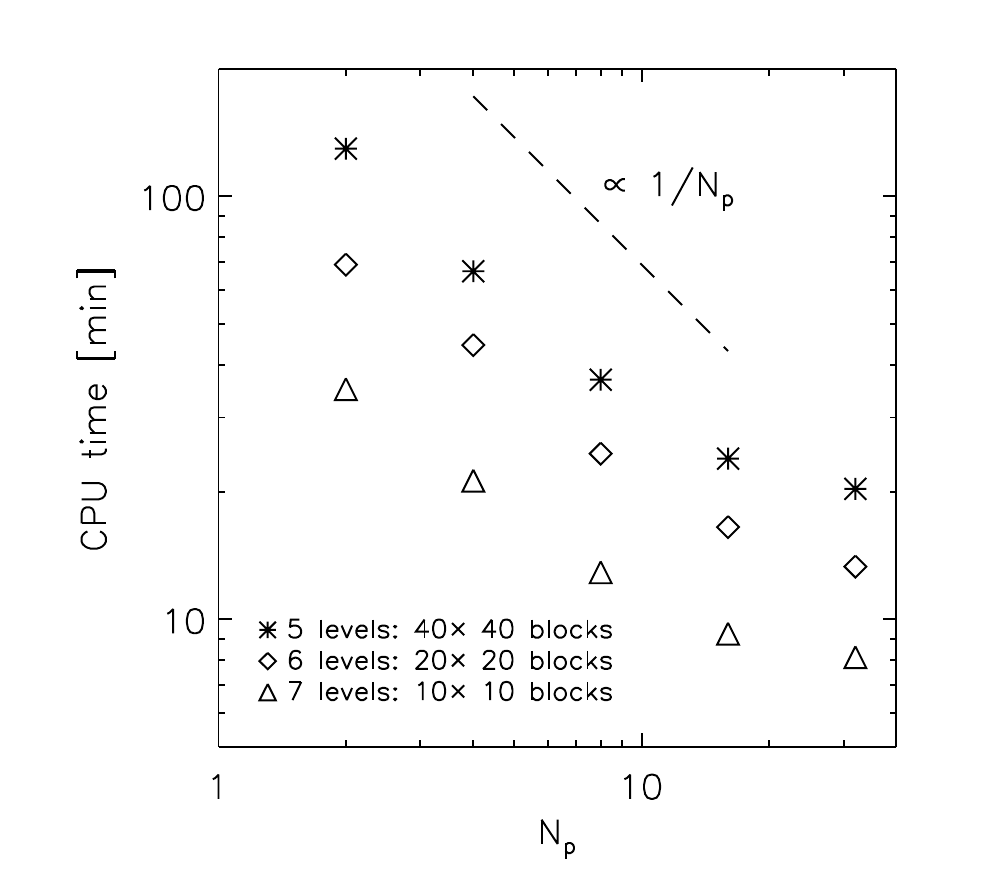}
\caption{Scaling of the {\sc Walicxe} code for the double Mach
  reflection test at an equivalent resolution of $2560\times 640$
  cells, on 2, 4, 8, 16, and 32 processors. The different symbols are
  for a different combination of levels of refinement and block size,
  as indicated in the legend.  We have included ({\it dashed} line)
  the ideal $\propto 1/N_p$ scaling for reference.}  
\label{fig:scaling}
\end{figure}

Adaptive mesh codes use a significant amount of CPU time to maintain
the AMR structure, and neighboring blocks can be in a different
processors at different times. This causes AMR codes to scale less than
ideally, but the CPU time and memory requirements can remain significantly
lower than for fixed grid codes.
Also, there are many variables that can impact the performance, and
many of them are problem dependent.
To illustrate this, we took the double Mach reflection test with an
equivalent resolution of $2560\times 640$ cells and ran it on
$2$, $4$, $8$, $16,$ and $32$ processors. To achieve the same
resolution at the highest level of refinement we used three
combinations: $5$ levels of refinement with $40\times 40$ cells
blocks, $6$ levels with blocks of $20\times 20$ cells, and $7$ levels with
$10\times 10$ blocks, the scaling results are presented in Figure
\ref{fig:scaling}.

We see from the figure a good performance of the code, almost linear
for a small number of processors. As we increase the number of
processors the portion of domain that is solved by each processors
becomes too small and the communication time becomes important
compared to the time to solve each region. 
Certainly, if we require an even finer resolution, the transition from
ideal, to not-so-ideal would happen at a larger number of processors.

Is is quite noticeable that by allowing the mesh to be finer the CPU
time drops dramatically. We must note that the outcome is particularly
good because the problem is well suited for an adaptive mesh, the
regions of sharp changes are well localized and are a small portion of
the entire computational domain. If refinement is required in a larger
portion of the domain, larger blocks and a small number of levels
would be beneficial.

\subsection{Fixing the kinked Mach stem (``carbuncle'') problem}
\label{ap:carb}
\begin{figure*}
\centering
\includegraphics[width=\textwidth]{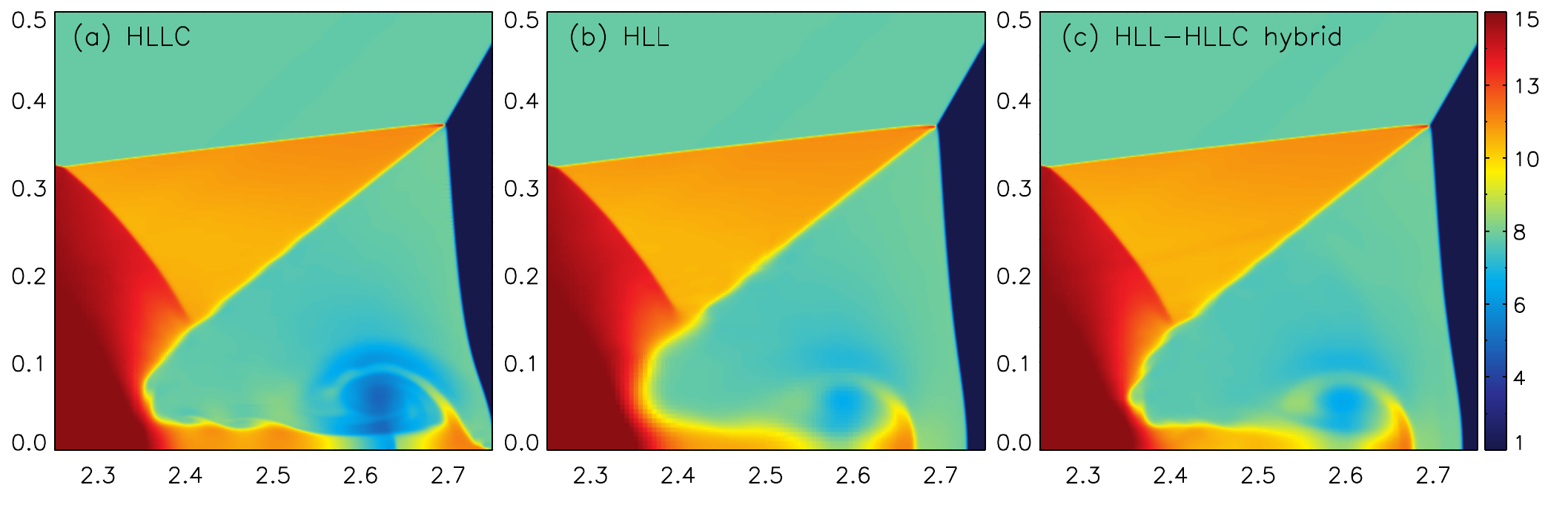}
\caption{Density maps for the $[2.25,2.75]\times[0,0.5]$ region of the double Mach reflection test with the ({\it a}) HLLC, ({\it b}) the HLL,  and ({\it c}) the hybrid HLL-HLLC  Riemann solvers. The density values are indicated by the bar at the right. All the runs were done with a maximum resolution of $\Delta x=\Delta y = 1/(640)$.}
\label{fig:HLL-HLLC}
\end{figure*}

The so-called numerical shock instability, or ``carbuncle'' problem was first reported by \citet{1994IJNMF..18..555Q}, and although a full understanding of the instability and its cure is not reached \citep{2000JCoPh.160..623L,  2008JCoPh.227.2560N}, researchers seem to agree that the culprit is insufficient viscosity when shocks align with the computational cell interfaces. Also, Riemann solvers that perform exceptionally well on shear layer problems (such as the Roe solver, \citealt{1981JCoPh..43..357R}) suffer more of this pathology, while more diffusive ones, such as the HLL solver have been considered virtually carbuncle-free.
In recent years, several remedies have been put forward to solve this instability, most of them for the solver of Roe \citep[see for instance ][]{2008JCoPh.227.2560N}. Among the remedies, a family of methods is to produce hybrid fluxes, combining a solver that is more diffusive, such as the HLL, with a more accurate, yet carbuncle-prone solver. In particular \citep{2009JCoPh.228.7634K, 2010IJNMFhuang}, have proposed a combination of the HLL and HLLC fluxes. We have implemented a similar hybrid flux formed by a linear combination of the two fluxes
\begin{equation}
\mathbf{F}_{\mathrm{HLL-HLLC}}=\beta_1 \mathbf{F}_{\mathrm{HLLC}}+ \beta_2 \mathbf{F}_{\mathrm{HLL}},
\end{equation}
where the $\beta_1+\beta_2=1$,  these coefficients are chosen in order to give more weight to the HLLC solver when shear flows are present, and to the HLL in the presence of strong shocks aligned with the grid. To do this we check the direction of the velocity difference vector between the left and right states: $\mathbf{\Delta q}=(u_R-u_L, v_R-v_L )$. In particular we use the magnitude of the projection of the unit vector  in the direction of $\mathbf{\Delta q}$ with the unit vector normal to the cell interface, that is : $\alpha_1=\vert u_R-u_L\vert / [(u_R-u_L)^2 +(v_R-v_L)^2]^{1/2}$. After some tests, we choose $\beta_1=\sqrt{\alpha}_1$, in figure \ref{fig:HLL-HLLC} we compare of the results from the double Mach reflection test with the HLLC, HLL, and HLL-HLLC solvers, all at $t=0.2$, and all with the same equivalent resolution $\Delta x=\Delta y = 1/(640)$.

We can see from the Figure (panel {\it a}) how the HLLC solver alone produces a kinked Mach stem, but resolves sharply the contact discontinuity (the density change at an angle of $~40\deg$. In fact some Kelvin-Helmholtz ``rolls'' start to appear, along this discontinuity and extend to the jet propagating along the reflecting surface ($x$-axis). For the HLL solver (panel {\it b}) the kinked Mach stem is controlled (barely noticeable), but the diffusion at the contact surfaces is evident, producing a smeared solution compared to the more accurate HLLC solver. In panel ({\it c}) we show the results with the hybrid fluxes, and the best of the other two solvers is evident, a barely noticeable kinked Mach stem, with a sharp contact surface.

\bibliography{Master}

\end{document}